\documentclass[onecolumn,preprintnumbers,nofootinbib,pagenumbers]{revtex4-1}

\usepackage{graphicx}
\usepackage[colorlinks=true,urlcolor=blue]{hyperref}
\usepackage{amsfonts}
\usepackage{amssymb}
\usepackage{amsmath}
\usepackage{mathrsfs}
\usepackage{slashed}

\newcommand{\Tr}{\mathop{\mathrm{Tr}}}
\DeclareMathOperator{\sgn}{sgn}
\usepackage{ifpdf}

\begin{document}

\title{An analysis of a minimal vectorlike extension\\ of the Standard
Model}

\author{V.\,Beylin{$^1$}\footnote{E-mail: vitbeylin@gmail.com}, M.\,Bezuglov{$^1$}\footnote{E-mail: bezuglov.ma@phystech.edu},
V.\,Kuksa{$^1$}\footnote{E-mail: vkuksa47@mail.ru},
N.\,Volchanskiy{$^{1,2}$}\footnote{E-mail: nikolay.volchanskiy@gmail.com}}

\affiliation{{$^1$}Research Institute of Physics, Southern Federal
University, 344090 Rostov-on-Don, Pr. Stachky 194, Russian
Federation}

\affiliation{{$^2$}Bogoliubov Laboratory of Theoretical Physics,\\Joint Institute of Nuclear Research, 141980 Dubna, Russia}

\begin{abstract}
We analyze an extension of the Standard Model with an additional
$SU(2)$ hypercolor gauge group keeping the Higgs boson as a
fundamental field. Vectorlike interactions of new hyperquarks with
the intermediate vector bosons are explicitly constructed. We also
consider pseudo-Nambu--Goldstone bosons caused by the symmetry
breaking $SU(4)\to Sp(4)$. A specific global symmetry of the model
with zero hypercharge of the hyperquark doublets ensures the
stability of a neutral pseudoscalar field. Some possible
manifestations of the lightest states at colliders are also
examined.
\end{abstract}


\maketitle

\section{Introduction}
The experimental detection of the Higgs boson \cite{1H,2H} with mass
$M_{H}\approx 125\,\mbox{GeV}$ leaves unanswered many questions
of the Standard Model (SM) (see \cite{Kaz}, for
example). A part of the SM puzzles can be solved by supersymmetry
(SUSY) \cite{SU1,SU2}. Unfortunately, there are no any clear
indications that SUSY manifests itself in the experiments near
a ``naturalness'' scale $\sim 1 \, \mbox{TeV}$. Obviously, SUSY is
not rejected at all, but sparticles and their interactions are now
expected to be observed at a much higher scale, $\sim 5\text{--}10\,
\mbox{TeV}$, because the parameter space of SUSY models is
increasingly constrained by the LHC data \cite{SU3,SU4,SU5}.

Besides SUSY, a lot of ways are proposed to enlarge SM: an addition of
extra $U(1)$ groups, multi-Higgs and technicolor
(TC) models, and many others (see reviews \cite{Morrissey20121,
Kaz} and references therein). However, we currently have not found
any comprehensive variant of the theory of ``everything''
(excepting, possibly, string theory which has no phenomenological
applications for now), so all problems of SM cannot be solved
simultaneously. An origin of Dark Matter (DM) is also one of the
known SM problems. At the moment we are skeptical of any
manifestations of (sufficiently light) neutralino as the DM particle
\cite{SUDM}. Note, there are a lot of other DM candidates which are
suggested and discussed \cite{HDM1, HDM2, HDM3, TCDM1, TCDM2, TCDM3,
TCDM4, TCDM5, TCDM6}. For example, DM can originate from the Higgs
sector too (e.g., the inert Higgs model) \cite{Do, HHI}.

From a ``technical'' viewpoint, technicolor scenario
\cite{TC1,TC2,TC3,ETC} means a ``duplication'' of an analog of the
QCD sector at a higher energy scale with confinement of the extra
techni-fermions and techni-gluons. Originally, TC models were
suggested to introduce dynamical electroweak (EW) symmetry breaking
(EWSB) without fundamental Higgs scalars. Corresponding scalar boson
arises in this case as a bound state of techni-quarks---these models
are higgsless (note also so called ``see-saw'' mechanism giving a
light scalar boson in TC) \cite{TC4,TC5,TC6,TC7,TC71,TC8}. In this
way both structure and interactions of the T-strong confined sector
are considered as extra options to solve some SM problems (see Refs.
\cite{TC9,TC10,TC11,TC12ou,Sundrum}). It seems that the
discovery of the Higgs boson closes some higgsless technicolor scenarios and many investigations concentrate now on extra fermion sectors in
confinement (the so-called hypercolor models) as a source of
composite states and Dark Matter candidates.

Contributions of additional fields to the SM precision
parameters are crucial for the models---variety of them is
constrained \cite{TC4} by the experimentally required values of
Peskin--Takeuchi (PT) parameters \cite{PT1,PT2,PT3,PT4,PT5,TC12ou}.
So, to select a realistic and reasonable extension, it is necessary
to calculate EW polarization operators with an account of the model
contributions. Then, the comparison of calculated values of
$S$, $T$, $U$ parameters with the experimental data gives some constraint on the structure of the model.
As a rule, in the models with chirally non-symmetric fermions
there appear unacceptable contributions to the PT parameters.
It is the main reason why vectorlike models has been under
consideration recently \cite{TC12ou,Sundrum,RevTC,DMou1,PEPou}.

Thus, multiplet and chiral structure of the new fermion sector is a
principal characteristic of SM extension. In the framework of
technicolor models, as a rule, such multiplets have a standard-like
$SU(2)_L$ structure, namely left-hand doublets and right-hand
singlets \cite{San08,PT6}. In the hypercolor models,
chirally-symmetric (with respect to the weak group) set of new
fermions is used \cite{Kilic2010Vectorlike}. However, this
chirally-symmetric fermion sector
 crucially differs from the
standard one, so interpretation of the gauge fields as standard vector bosons is hypothetical.

In this work, we suggest a construction of vectorlike weak
interaction which starts from standard-like chirally non-symmetric
set of new fermions doublets. This program has been carried out for
zero hypercharge in the simplest model with two hyperquark
(H-quark) generations and two hypercolors (HC), $N_{HC}=2$
\cite{PEPou,Pasechnik2016Composite}. We consider this scenario for
the case of non-zero hypercharge and show that two left doublets of
H-quarks can be transformed into one doublet of Dirac H-quarks with
vectorlike weak interaction. This possibility can be realized if the
hypercharges of generations have the same value and opposite signs.
Importantly, this condition is in accordance with the absence of
anomalies in the model. To form the Dirac states which correspond to
constituent quarks, we have used a scalar field having non-zero
vacuum expectation value (v.e.v.). This field is introduced as
a scalar singlet pseudo-Nambu--Goldstone (pNG) boson in the framework of the simplest
linear sigma-model. We consider in detail the structure of the
pNG multiplet which is defined by the global
symmetry breaking $SU(4)\to Sp(4)$. It is also shown that the
Lagrangian of this minimal extension has a specific global
symmetries making neutral H-baryon and H-pion states stable.

The paper is organized as follows. In the second section, we
construct vectorlike interactions for the case of $SU(2)$ H-color
and EW groups with even generations. The total Lagrangian together with the pNG bosons
 is considered in the third section.
The principal part of the physical Lagrangian of the model is
presented in the fourth section, where we demonstrate the presence of a specific
discrete symmetry that leads to the stability of a
pseudoscalar state. In the fifth section, we analyze the main
phenomenological consequences of the model.

\section{\label{sec:2}Vectorlike interaction of the gauge bosons with H-quarks}

An essential point is the choice of chiral structure of the H-quark
multiplets. It is known that chirally non-symmetric interaction of
the extra fermions with the SM bosons may contradict to restrictions
on Peskin--Takeuchi parameters. Thus, it is reasonable to consider
vectorlike (chiral-symmetric) interaction of (initially
standard-like) H-quarks with $\mathit{Z}$ and $\mathit{W}$-bosons.
We construct such interactions explicitely for the case of even
generations of two-color $(N_{HC}=2)$ H-quarks.

In the simplest scenario with two generations ($A=1,\,2$) of
left-handed H-quarks, the bi-doublet of these quarks is presented as
a matrix $Q^{a\underline a}_{L(A)}$, where $a=1, \,2$ and $\underline a =1, \,2$
are indices of $SU(2)_L$ and $SU(2)_{HC}$ fundamental
representations respectively. (In the following all indices related to the hypercolor group are underlined.)

This bi-doublet transforms under $U(1)_Y\otimes SU(2)_L\otimes
SU(2)_{HC}$ as
 \begin{equation}\label{2.1}
(Q^{a\underline a}_{L(A)})^{'}=Q^{a\underline a}_{L(A)}+ig_B Y_A\theta
Q^{a\underline a}_{L(A)}+\frac{i}{2}g_W\theta_k
\tau_k^{ab}Q^{b\underline a}_{L(A)}+\frac{i}{2}g_{HC}\varphi_{\underline k}\tau_{\underline k}^{\underline a \underline b}
Q^{a\underline b}_{L(A)}.
\end{equation}
Here $Q^{1\underline a}_{L(A)}=U^{\underline a}_{L(A)}$,
$Q^{2\underline a}_{L(A)}=D^{\underline a}_{L(A)}$ and the H-quarks charges
$q_{U,D}$ are defined by the arbitrary hypercharges $Y_A$. The
right-handed singlets (with respect to electroweak $SU(2)_L$ group)
have the following group transformations:
\begin{equation}\label{2.2}
(S^{\underline a}_{R(A)})^{'}=S^{\underline a}_{R(A)}+ig_B Y_{R(A)} \theta
S^{\underline a}_{R(A)}+\frac{i}{2}g_{HC}\varphi_{\underline k}\tau_{\underline k}^{\underline a \underline b}S^{\underline b}_{R(A)},
\end{equation}
where $A=1,2$ and $Y_{R(A)}$ are hypercharges of singlets. Now, the
charge conjugation operation, $\mathbf{\hat{C}}$, is applied to the
fields of the second generation keeping the first generation of
H-quarks unchanged:
\begin{equation}\label{2.3}
Q^{Ca\underline a}_{L(2)}=\mathbf{\hat{C}}Q^{a\underline a}_{L(2)}.
\end{equation}
The transformation properties of the charge conjugated fields have
the form
\begin{equation}\label{2.4}
(Q^{Ca\underline a}_{L(2)})^{'}=Q^{Ca\underline a}_{L(2)}-ig_B Y_2
Q^{Ca\underline a}_{L(2)}-\frac{i}{2}g_W\theta_k
(\tau_k^{ab})^*Q^{Cb\underline a}_{L(2)}-\frac{i}{2}g_{HC}\varphi_{\underline k}(\tau_{\underline k}^{\underline a \underline b})^*
Q^{Ca\underline b}_{L(2)}.
\end{equation}
Then, we redefine the H-quark fields (the fermion chirality is
changed by the charge conjugation):
\begin{equation}\label{2.5}
Q^{a\underline a}_{R(2)}=\epsilon^{ab}\epsilon^{\underline a \underline b}Q^{Cb\underline b}_{L(2)},\qquad \epsilon^{ab}=\begin{pmatrix}0&1\\-1&0\end{pmatrix}.
\end{equation}

Further, we multiply both sides of (\ref{2.4}) by
$\epsilon^{ab}\epsilon^{\underline a \underline b}$ and use the following
properties of $SU(2)$ group matrices:
\begin{equation}\label{2.6}
\epsilon^{ac} \epsilon^{bc}=\delta^{ab},\,\,\,\epsilon^{ab}(\tau^{bc}_k)^*\epsilon^{cf}=\tau_k^{af}.
\end{equation}

Using the redefinition (\ref{2.5}), from (\ref{2.4}) we get:
\begin{equation}\label{2.7}
(Q^{a\underline a}_{R(2)})^{'}=Q^{a\underline a}_{R(2)}-ig_B Y_2
Q^{a\underline a}_{R(2)}+\frac{i}{2}g_W\theta_k\tau_k^{ab}Q^{b\underline a}_{R(2)}+
\frac{i}{2}g_{HC}\varphi_{\underline k}\tau_{\underline k}^{\underline a \underline b}Q^{a\underline b}_{R(2)}.
\end{equation}
This transformation law coincides with the one given by the formula
(\ref{2.1}) for the first generation $(A=1)$ when $Y_2=-Y_1$.

Thus, we have constructed the right-handed field partner of the first
generation, using the second generation of the left-handed fields in
two steps: charge conjugation and redefinition. Therefore, composing
these fields we have a Dirac state:
\begin{equation}\label{2.8}
Q^{a\underline a}=Q^{a\underline a}_{L(1)}+Q^{a\underline a}_{R(2)}=Q^{a\underline a}_{L(1)}+\epsilon^{ab}\epsilon^{\underline a \underline b}Q^{Cb\underline b}_{L(2)}.
\end{equation}
Because both parts (left- and right-handed) of the field have the same
transformation properties, namely (\ref{2.1}), then the Dirac
H-quarks interact with the EW vector bosons as chiral symmetric
fields.

Analogously, the right-handed field $S^{\underline a}_{R(2)}$ is redefined
as follows:
\begin{equation}\label{2.9}
S^{\underline a}_{L}=\epsilon^{\underline a \underline b}\mathbf{\hat{C}}S^{\underline b}_{R(2)}.
\end{equation}
 This redefined field transforms as the right-handed singlet $S_{R(1)}$ if $Y_{R(2)}=-Y_{R(1)}$ in full analogy with the previous case.
This representation of the H-fields allows us to get a usual Dirac
mass term after the summation of left and right parts. Both current
and constituent H-quark masses can be introduced because the mass
term does not violate the model symmetry. The simplest way to do
this is to use a singlet real scalar, $s$, which has a non-zero
v.e.v., $s=\tilde{\sigma} +u$, where $u=\langle s\rangle$. Just
interaction of the H-quarks with this scalar field provides Dirac
type mass term for H-quarks. Note, to get a Dirac state with the
vectorlike interaction from two Weyl spinors, we should require the
initial fields for the first and second families to have opposite
hypercharges, $Y_1 =-Y_2$. The same requirement follows from the
condition of the absence of anomalies in the model. It should be
noted that the suggested construction of vectorlike interaction is
valid due to unique properties of $SU(2)_{HC}$ group and for the
case of an even number of generations.

The gauge part of the model Lagrangian directly follows from
(\ref{2.1}) and (\ref{2.2}):
\begin{align}\label{2.10}
L(Q,S)={}&-\frac{1}{4}T^{\underline k}_{\mu\nu}T^{\mu\nu}_{\underline k}
+i\bar{Q}\gamma^{\mu}(\partial_{\mu}-ig_B
Y_1 B_{\mu}-\frac{i}{2}g_W W^k_{\mu}\tau_k-
\frac{i}{2}g_{HC}T^{\underline k}_{\mu}\tau_{\underline k})Q-m_Q\bar{Q}Q\notag\\
&+i\bar{S}\gamma^{\mu}(\partial_{\mu}-ig_B Y_{R(1)}
B_{\mu}-\frac{i}{2}g_{HC}T^{\underline k}_{\mu}\tau_{\underline k})S-m_S\bar{S}S,
\end{align}
where $T_{\mu}^{\underline k}$ is a H-gluon field. The mass
terms are formally included in (\ref{2.10}) because they do not
break $SU(2)_{HC}$-symmetry of the model. The status of the
$SU(2)_L$-singlet H-quark significantly differs from that of the
standard quarks. The standard quark singlet is a right-handed part
of the Dirac fermion state, while $\mathit{S}$-quark consists of the
two initial chiral singlets. It should be noted that the singlet $S$
can be useful since a composite H-meson $\bar{Q}S$ is a
representation of the groups $U(1)_Y\otimes SU(2)_L$. The standard
Higgs doublet is the same representation, that is, the Higgs field
can be considered as a composite state of the singlet and doublet
H-quarks. However, due to the fields $Q$ and $S$ are independent,
from now on, the $SU(2)_L$ singlet states can be not included into the
consideration.

\section{Fundamental Higgs boson, two-color fermions, and pseudo-Nambu--Goldstone bosons in the linear sigma model}

Here, we construct a linear sigma model involving the constituent H-quarks and lowest pseudo(scalar)
H-hadrons---$\sigma$ H-meson, pNG states, and their opposite-parity partners \cite{San08,PT6,PT7,PT8,Vysotskii1985Spontaneous}. As it was shown
in \cite{Vysotskii1985Spontaneous,Pauli1957Conservation,Gursey1958Relation} (see also more recent papers \cite{Lewis2012Light,Cacciapaglia2014Fundamental}), the Lagrangian
(\ref{2.10}) in the limit $m_{Q}\to 0$, $g_W\to 0$ has a global $SU(4)$ symmetry
corresponding to rotations in the space of the four initial chiral fermion fields. The Lagrangian with
non-zero $m_Q$ can be rewritten in the form which explicitly reveals
the violation of symmetry $SU(4)\to Sp(4)$ by the mass term
\cite{Lewis2012Light,Cacciapaglia2014Fundamental}. For $m_Q=0$ the
Lagrangian retains the full $SU(4)$ symmetry but, in an analogy with QCD, one might expect the dynamical symmetry breaking by vacuum
expectation value $\langle \bar{U}U+\bar{D}D\rangle \ne 0$. This
v.e.v.\ has the mass term structure and leads to the dynamical
breaking of the symmetry $SU(4)\to Sp(4)$. As a result, the broken
generators of $SU(4)$ would be accompanied by a set of pNG states.
The spectrum of the pNG states depends on the way of symmetry
breaking.

The global symmetry of two-color QCD with $N_{\tilde F}$ Dirac
quarks in the limit of zero masses is $SU(2 N_{\tilde F}$), with the
chiral group being its subgroup, ${SU}(N_{\tilde F})_L \otimes
{SU}(N_{\tilde F})_R \subset {SU}(2 N_{\tilde F})$\footnote{This
statement is valid for any symplectic gauge theory
\cite{Barnard2014UV}. The group $SU(2)$ is isomorphic to the group
$Sp(2)$.} \cite{Pauli1957Conservation,*Gursey1958Relation}. This
global symmetry is often called  the Pauli--G\"{u}rsey symmetry.  The
quark condensate breaks the Pauli--G\"{u}rsey symmetry to its
subgroup $Sp(2 N_{\tilde F}$)
\cite{Vysotskii1985Spontaneous,Verbaarschot2004Supersymmetric}. In
the following we will consider the simplest case of two flavors
$N_{\tilde F}=2$.

We have only two possibilities to assign EW quantum
numbers to the two fundamental fermion constituents\footnote{For the
general case a classification of physically relevant ultraviolet completions
of composite Higgs models based on the coset $SU(4)/Sp(4)$ is given in
Ref.~\cite{Ferretti2014Fermionic, Barnard2014UV}, which consider
different gauge groups with arbitrary numbers of flavors and colors,
$ N_{\tilde F}$ and $N_{HC}$.}. These possibilities are
determined by the cancellation of gauge anomalies.
\begin{itemize}
\item \textit{V-A ultraviolet completion}. We can introduce a left-handed weak doublet
$Q_L = \left(\begin{smallmatrix} U_L \\ D_L \end{smallmatrix}\right)$ and two right-handed weak
singlets $U_R$ and $D_R$ with opposite hypercharges $Y(U_R) = - Y(D_R)$. It is the case that is
considered in most papers dealing with a new two-flavor confined sector
\cite{Simmons1989Phenomenology, Appelquist1999Enhanced, Duan2000Enhanced, Ryttov2008Ultra, Evans2010Minimal, Cacciapaglia2014Fundamental}.
\item \textit{Vectorlike ultraviolet completion}. Both left- and right-handed fermions are grouped as fundamental
representations of the weak $SU(2)_L$ group, $Q_L = \left(\begin{smallmatrix} U_L \\ D_L\end{smallmatrix}\right)$
and $Q_R = \left(\begin{smallmatrix} U_R \\ D_R\end{smallmatrix}\right)$  \cite{Pasechnik2016Composite, Beylin2016Model}.
The hypercharges of the doublets should be the same, $Y(Q_L) = Y(Q_R)$. In this case the Dirac mass term, $\bar Q_L Q_R + \bar Q_R Q_L$, is permitted by the EW symmetry.
\end{itemize}
In this paper, we study the case of the vectorlike ultraviolet
completion with zero hypercharges of the doublets.

At the fundamental level, the Lagrangian of two-flavor and two-color QCD \eqref{2.10} can be written in terms of a
left-handed quartet field:
\begin{gather}\label{eq:LP}
    L = -\frac{1}{4}T^{\underline k}_{\mu\nu}T^{\mu\nu}_{\underline k}
 + i \bar P_{L}^{\underline a} \slashed{D}_{\underline a \underline b} P_{L}^{\underline b} -
 \frac12 m_Q \left( \bar P_L^{\underline a} M_0 P_R^{\underline a} + \bar P_R^{\underline a} M_0^\dagger P_L^{\underline a} \right),
    \\ \label{eq:DP}
    D^\mu_{\underline a \underline b} = \partial^\mu \delta_{\underline a \underline b} -
    \frac{i}2 g_{HC} T^\mu_{\underline k} \tau^{\underline k}_{\underline a \underline b}  -\sqrt2 i g_W W^\mu_k \Sigma_k \delta_{\underline a \underline b},
\end{gather}
where
\begin{gather}
    P_{L}^{\underline a} = \begin{pmatrix} Q_{L(1)}^{\underline a} \\ Q_{L(2)}^{\underline a} \end{pmatrix},
    \qquad
    P_{R}^{\underline a} = \epsilon^{\underline a \underline b} (P_L^{\underline b} )^C
\end{gather}
are left- and right-handed quartet fields ($Q_{L(1)}$ and $Q_{L(2)}$ are left-handed doublets introduced in the previous Section).
The EW term in the covariant derivative \eqref{eq:DP} involves the matrices
\begin{gather}\label{eq:Sigma123}
    \Sigma_k = \frac1{2\sqrt2} \begin{pmatrix} \tau_k & 0 \\ 0 & \tau_k \end{pmatrix},
    \quad   k=1,\,2,\,3,
\end{gather}
that are three of ten $Sp(4)$ generators $\Sigma_\alpha$ satisfying the following conditions:
\begin{gather}\label{eq:sp4}
    \Tr \Sigma_\alpha =0, \qquad \Sigma_\alpha^\dagger = \Sigma_\alpha, \qquad \Tr \Sigma_\alpha \Sigma_\beta = \frac12 \delta_{\alpha\beta}, \qquad
    \Sigma_\alpha^T M_0 + M_0 \Sigma_\alpha= 0,
    \qquad
    \alpha,\,\beta = 1,\,2,\dots 10.
\end{gather}
The mass term in the Lagrangian \eqref{eq:LP} introduces the antisymmetric $4\times4$ matrix
\begin{gather}
 M_0 = -M_0^T = \begin{pmatrix} 0 & \epsilon \\ \epsilon & 0 \end{pmatrix}.
\end{gather}
We have used the matrix $M_0$ also to define the algebra of the $Sp(4)$ generators. Although $M_0$ has a
noncanonical form, it can be brought into the form $\left(\begin{smallmatrix} 0 & 1 \\ -1 & 0 \end{smallmatrix}
\right)$ or $\left(\begin{smallmatrix} \epsilon & 0 \\ 0 &
\epsilon \end{smallmatrix} \right)$ by a unitary
transformation.

The equivalence of the Lagrangians \eqref{2.10} and \eqref{eq:LP} was proved in the previous Section. It should be noted that the similar
rearrangement of the Lagrangian in terms of the left-handed fields would be possible in any sort of techni- or hyperchromodynamics with T/H-quarks
in selfcontragredient representation of T/H-confinement group. The fundamental representation of $SU(2)_{HC}$, which is symplectic and
pseudoreal representation, is just the simplest case. An aspect of this property is that the global symmetry group of the massless theory is larger than the chiral symmetry.

In the limit of vanishing $m_Q$ and $g_W$ the global symmetry group of the Lagrangian \eqref{eq:LP} is the Pauli--G\"{u}rsey group $SU(4)$
\cite{Pauli1957Conservation,Gursey1958Relation}, the chiral symmetry being a subgroup of the Pauli--G\"{u}rsey group:
\begin{gather}\label{eq:LTQ}
    P_L^{\underline a} \to U P_{L}^{\underline a}, \qquad  P_R^{\underline a} \to U^* P_{R}^{\underline a},
    \qquad U \in SU(4).
\end{gather}
The mass term of the current H-quarks breaks the group $SU(4)$ explicitly. Indeed, if we consider infinitesimal transformations
$U=1+i \theta_\alpha \Sigma_\alpha$, $\theta_\alpha \ll 1$, it is readily seen that the mass term in the Lagrangian \eqref{eq:LP} is left
invariant by the generators satisfying the conditions \eqref{eq:sp4}, that is the mass term is invariant under the subgroup $Sp(4)$ of the
Pauli--G\"{u}rsey group (see \cite{Lewis2012Light,Cacciapaglia2014Fundamental}). H-quark condensate $\langle \bar QQ \rangle$ has the same
spinor structure as the mass term. Thus, the dynamical breaking by the condensate $\langle \bar QQ \rangle$ should be also $SU(4)\to Sp(4)$
\cite{Vysotskii1985Spontaneous,Verbaarschot2004Supersymmetric}. If the current H-quark masses are significantly smaller than the scale of the
spontaneous breaking of the Pauli--G\"{u}rsey group, we have the situation similar to the one in well-established QCD of light quarks.
Putting it in terms natural to the quark-meson sigma models, there are five pNG bosons associated with the five ``broken'' generators of the
group $SU(4)$, these bosons acquire small masses due to the small explicit breaking of the global symmetry of the model, while the constituent
masses of the H-quarks are generated mostly by the dynamical symmetry breaking.

Before leaving our consideration of the Lagrangian of the fundamental current H-quarks, we should note that apart from the Pauli--G\"{u}rsey
group $SU(4)$ the Lagrangian \eqref{eq:LP} possesses an additional global $U(1)$ symmetry as well as a new discrete symmetry. The former symmetry
leads to conservation of an analog of the baryon number, while the latter one is a generalization of the $G$-parity of QCD. The important
consequences of these symmetries are discussed at the end of this Section and in the next one.

Now, we proceed to construct an effective Lagrangian of a linear quark-hadron sigma model $SU(4) \cong SO(6) \to SO(5) \cong Sp(4)$.
This model describes the interactions of the constituent H-quarks and lightest (pseudo)scalar H-hadrons. The Lagrangian of the H-quark sector of the model reads
\begin{gather}\label{eq:ctq}
    {L} = i \bar P_{L} \slashed D P_{L} -\sqrt2 \kappa \left( \bar P_{L} M P_{R} + \bar P_R M^\dagger P_L \right),
    \\ \label{eq:DP_constituent}
    D_\mu P_L = \partial_\mu P_L  -\sqrt2 i g_W W_\mu^k \Sigma_k P_L.
\end{gather}
Here $\kappa$ is a H-quark--H-hadron coupling constant. The matrix $M$ of spin-0 H-hadrons is antisymmetric. Its transformation law under the global symmetry $SU(4)$ is
\begin{gather}\label{eq:M_transf}
    M \to UMU^T, \qquad U \in SU(4) .
\end{gather}
Being a complex antisymmetric matrix with 12 independent components, the field $M$ can be conveniently expanded in terms of five
``broken'' generators $\beta_{\dot\alpha}$ of the Pauli--G\"{u}rsey group:
\begin{gather}
    M = \left[ \frac1{2\sqrt2} (A_0+iB_0) + (A_{\dot\alpha}+iB_{\dot\alpha}) \beta_{\dot\alpha} \right] M_0.
\end{gather}
The generators $\beta_{\dot\alpha}$ are subjected to the conditions
\begin{gather}
    \Tr \beta_{\dot\alpha} =0, \qquad \beta_{\dot\alpha}^\dagger = \beta_{\dot\alpha},
    \qquad \Tr \beta_{\dot\alpha} \beta_{\dot\gamma} = \frac12 \delta_{\dot\alpha\dot\gamma},
    \qquad \Tr \Sigma_\alpha \beta_{\dot\alpha} = 0,
    \\
    \qquad \beta_{\dot\alpha}^T M_0 - M_0 \beta_{\dot\alpha}=0,
    \qquad
    \dot\alpha,\,\dot\gamma=1,\,2,\dots 5, \qquad \alpha = 1,\,2,\dots 10
\end{gather}
and can be written explicitly as
\begin{gather}
    \beta_k = \frac1{2\sqrt2} \begin{pmatrix} \tau_k & 0 \\ 0 & -\tau_k \end{pmatrix}, \qquad k=1,\,2,\,3,
    \qquad
    \beta_4 = \frac1{2\sqrt2} \begin{pmatrix} 0 & 1 \\ 1 & 0 \end{pmatrix},
    \qquad
    \beta_5 = \frac{i}{2\sqrt2} \begin{pmatrix} 0 & 1 \\ -1 & 0 \end{pmatrix}.
\end{gather}
Now the Lagrangian of constituent H-quarks \eqref{eq:ctq} can be put
into the following form:
\begin{gather}
    {L} = i \bar Q \slashed{D} Q - \kappa u \bar Q Q
    \notag\\\label{eq:lp}
    -\kappa \left[ \sigma' \bar Q Q + i \tilde{\eta} \bar Q \gamma_5 Q
     + \tilde{a}_k \bar Q \tau_k Q + i \tilde{\pi}_k \bar Q \gamma_5 \tau_k Q
    +\frac{1}{\sqrt2}  \left( A^0 \bar Q_{a\underline a} \epsilon_{ab} \epsilon_{\underline a \underline b} Q_{b\underline b}{}^{C}
    +iB^0 \bar Q_{a\underline a} \epsilon_{ab} \epsilon_{\underline a \underline b} \gamma_5 Q_{b\underline b}{}^{C} + \text{h.c.} \right) \right],
\\
    \label{eq:DQ}
    D_\mu Q = \partial_\mu Q - \frac{i}2 g_W W_\mu^k \tau_k Q,
\end{gather}
where $\gamma_5=i\gamma^0\gamma^1\gamma^2\gamma^3$ and
\begin{gather}
    \sigma' = A_0 - u, \quad \tilde{\eta} = B_0,
\\
    \tilde{a}_k = A_k, \quad \tilde{\pi}_k = B_k, \quad
    k=1,\,2,\,3,
\\
    A^0 = \frac1{\sqrt2} ( A_4 + i A_5 ), \quad
    B^0 = \frac1{\sqrt2} ( B_4 + i B_5 ).
\end{gather}
From now on we use tildes to distinguish hypermesons from usual ones. The v.e.v.\ $u=\langle A_0 \rangle \sim \langle \bar Q Q \rangle$ breaks
the global symmetry $SU(4)$ spontaneously.

As it is seen from the form of the covariant derivative \eqref{eq:DP_constituent}, the local electroweak group is embedded into global $Sp(4)$ and
breaks it as well as its chiral subgroup explicitly. The covariant derivative of the (pseudo)scalars follows from the transformation properties of $M$:
\begin{gather}\label{eq:DM}
    D_\mu M = \partial_\mu M
         -\sqrt2 i g_W W_\mu^k (\Sigma_k M + M \Sigma_k^T ).
\end{gather}

Using the above derivative, the scalar sector of the model can be written as follows:
\begin{gather}
    L = D_\mu \mathcal{H}^\dagger \cdot D^\mu \mathcal{H} + \Tr D_\mu M^\dagger \cdot D^\mu M  - U
    \notag\\
    = \frac12 \left( D_\mu h \cdot D^\mu h +D_\mu h_k \cdot D^\mu h_k + \partial_\mu \tilde{\sigma} \cdot \partial^\mu \tilde{\sigma} + D_\mu \tilde{\pi}_k\cdot
    D^\mu \tilde{\pi}_k + \partial_\mu \tilde{\eta} \cdot \partial^\mu \tilde{\eta} + D_\mu \tilde{a}_k\cdot D^\mu \tilde{a}_k \right)
    \notag\\\label{eq:LM}
    + \partial_\mu \bar A^0\cdot \partial^\mu A^0 + \partial_\mu \bar B^0\cdot \partial^\mu B^0 - U,
\end{gather}
where the covariant derivatives of the H-meson fields read
\begin{gather}\label{eq:dpi}
    D_\mu \tilde{\pi}_k = \partial_\mu \tilde{\pi}_k + g_W e_{klm} W_\mu^l \tilde{\pi}_m ,
    \qquad
    D_\mu \tilde{a}_k = \partial_\mu \tilde{a}_k + g_W e_{klm} W_\mu^l \tilde{a}_m .
\end{gather}
In \eqref{eq:LM} it is assumed that the Higgs doublet $\mathcal H$ of SM is fundamental, not composite.
Its transformation properties are defined as usual in SM---the covariant derivative of $\mathcal H$ is
\begin{gather}
D_\mu \mathcal{H} =\partial_\mu \mathcal{H} +\frac{i}{2} g_B B_\mu \mathcal{H} -\frac{i}{2} g_W
W_\mu^k \tau_k \mathcal{H},
\end{gather}
or equivalently
\begin{gather}
   \mathcal{H} = \frac{1}{\sqrt2} \begin{pmatrix} h_2+ih_1 \\ h-ih_3\end{pmatrix}
    = \frac{1}{\sqrt2} \left( h + ih_k\tau_k\right) \begin{pmatrix} 0 \\ 1 \end{pmatrix},
    \\
    D_\mu h
        = \partial_\mu h
             +\frac12 (g_B \delta_3^k B_\mu +g_W W_\mu^k ) h_k,
\\
    D_\mu h_k
        =\partial_\mu h_k
             -\frac12 (g_B \delta_3^k B_\mu +g_W W_\mu^k ) h
             -\frac12 e_{klm} (g_B \delta_3^l B_\mu -g_W W_\mu^l ) h_m.
\end{gather}

In the Lagrangian \eqref{eq:LM} the potential term $U$ consists of self-interactions of the scalar fields:
\begin{gather}\label{eq:U}
    U = -\sum_{i=0}^3 \mu_i^2 I_i + \sum_{i \leq j=0}^3 \lambda_{ij} I_i I_j ,
\end{gather}
where $I_0$ is the $SU(2)_L \otimes U(1)_Y$ invariant of the SM Higgs doublet and $I_i$, $i=1,\,2,\,3$, are three independent $SU(4)$ invariants of the field $M$:
\begin{gather}
    I_0 = \mathcal{H}^\dagger \mathcal{H}  = \frac12 (v+h)^2,
    \\
    I_1 = \Tr M^\dagger M - 4 \mathop{\mathrm{Re}} \mathop{\mathrm{Pf}} M = \frac12 \left[ (u+\sigma')^2 + \tilde{\pi}_k \tilde{\pi}_k + 2 \bar B^0 B^0 \right],
    \\
    I_2 = \Tr M^\dagger M + 4 \mathop{\mathrm{Re}} \mathop{\mathrm{Pf}} M = \frac12 \left[ \tilde{\eta}^2 + \tilde{a}_k \tilde{a}_k + 2 \bar A^0 A^0\right],
    \\
    I_3 = 4 \mathop{\mathrm{Im}} \mathop{\mathrm{Pf}} M = -(u+\sigma') \tilde{\eta} + \tilde{a}_k \tilde{\pi}_k + \bar B^0 A^0 + \bar A^0 B^0.
\end{gather}
Here $\mathop{\mathrm{Pf}} M = -\frac14 \Tr M \tilde M = \frac18 \epsilon_{prst} M_{pr} M_{st}$ is the Pfaffian of $M$; $\epsilon_{prst} $ is
the 4-dimensional Levi-Civita symbol ($\epsilon_{1234}=+1$), $v=\langle h \rangle$ is the Higgs-field v.e.v. We consider only renormalizable
self-interactions of the scalar fields, although renormalizability in general has nothing to do with effective field theories. The invariant $I_3$ is
odd under $CP$ conjugation. $CP$ invariance implies that $\lambda_{03} = \lambda_{13} = \lambda_{23} = 0$.

Tadpole equations for $v,\,u \neq 0$:
\begin{gather}\label{eq:tadpoles}
    \mu_0^2 = \lambda_{00} v^2 + \frac12 \lambda_{01} u^2,
    \qquad
    \mu_1^2 = \lambda_{11} u^2 + \frac12 \lambda_{01} v^2 + \frac{\zeta \langle \bar Q Q \rangle}{u}.
\end{gather}
Vacuum stability is ensured by the following inequalities:
\begin{gather}
    \Lambda_{11} = \lambda_{11} - \frac{\zeta \langle \bar Q Q \rangle}{2u^3} > 0, \qquad \lambda_{00} > 0,
    \qquad
    4 \lambda_{00} \Lambda_{11} - \lambda_{01}^2 > 0.
    \label{eq:vacineq}
\end{gather}
Deriving \eqref{eq:tadpoles} and \eqref{eq:vacineq} we have taken
into account a tadpole-like source term $L_\text{SB} = -\zeta
\langle \bar Q Q \rangle (u+\sigma')$, where $\zeta$ is a parameter
proportional to the current mass $m_Q$ of the H-quarks. Such term in
phenomenological fashion communicates effects of explicit breaking
of the $SU(4)$ global symmetry to the vacuum parameters and the
H-hardon spectrum. This resembles QCD---the chiral symmetry is
broken both dynamically (with the quark condensate $\langle \bar q q
\rangle$ as an order parameter) and explicitly (by the quark
masses). In the sigma models with linear realization of the chiral
symmetry, the spontaneous breaking is induced by v.e.v.\ of $\sigma$
meson field. The effects of the explicit breaking can be mimicked by
different chirally non-invariant terms
\cite{1979PhRvC..19.1965C,2000PhRvC..61b5205D,1996NuPhA.603..239D},
but  the most common one, which is sometimes reffered to as
``standard breaking'', is a tadpole-like $\sigma$ term (see
\cite{1969RvMP...41..531G,2013PhRvD..87a4011P}, for example).

The masses of the (pseudo)scalar fields read
\begin{gather}
    m_{\tilde{\sigma},H}^2 = \lambda_{00} v^2 + \Lambda_{11} u^2 \pm \sqrt{(\lambda_{00} v^2 - \Lambda_{11} u^2)^2 + \lambda_{01}^2 v^2 u^2},
    \qquad m_{\tilde{\pi}}^2 = m_B^2 = -\frac{\zeta \langle \bar Q Q \rangle}{u},
    \\
    m_{\tilde{\eta}}^2 = m_{\tilde{a}}^2 + 2 \lambda_{33} u^2,
    \qquad m_{\tilde{a}}^2 = m_A^2 = -\mu_2^2 + \frac12 \lambda_{02} v^2 + \frac12 \lambda_{12} u^2 .
\end{gather}

The physical Higgs boson becomes partially composite receiving a tiny admixture of the scalar field $\sigma'$:
\begin{gather}
    h = \cos\theta_s H - \sin\theta_s \tilde{\sigma},
    \qquad
    \sigma' = \sin\theta_s H + \cos\theta_s \tilde{\sigma},
    \qquad
    \tan2\theta_s = \frac{\lambda_{01} v u }{\lambda_{00} v^2 - \Lambda_{11} u^2},
    \qquad
    \sgn \sin\theta_s = -\sgn \lambda_{01},
\end{gather}
where $h$ and $\sigma'$ are the fields being mixed, while $H$ and
$\tilde{\sigma}$ are physical ones.

Finally, the self-interactions of scalar fields take the form
\begin{gather}
    {L} = -\lambda_{00} h^3 \left( v + \frac14 h \right) -  \frac14 \lambda_{11} \left( B_{\dot\alpha} B_{\dot\alpha} + \sigma'{}^2 \right)
    \left( B_{\dot\alpha} B_{\dot\alpha} + \sigma'{}^2 + 4 u \sigma' \right)
    \notag\\
    - \frac14 \lambda_{01} h \left[ \left( 2 v + h \right) \left(B_{\dot\alpha} B_{\dot\alpha} + \sigma'{}^2 \right) + 2 u \sigma' h \right]
    - \frac14 \lambda_{02} h ( 2 v + h ) ( A_{\dot\alpha} A_{\dot\alpha} + \tilde{\eta}^2 )
    \notag\\
    - \frac14 \lambda_{12} \left( B_{\dot\alpha} B_{\dot\alpha} + \sigma'{}^2 +2 u \sigma' \right) ( A_{\dot\alpha} A_{\dot\alpha} + \tilde{\eta}^2 )
    - \frac14 \lambda_{22} ( A_{\dot\alpha} A_{\dot\alpha} + \tilde{\eta}^2 )^2
    \notag\\
    - \lambda_{33} \left[ -(u+\sigma') \tilde{\eta} + \tilde{a}_k \tilde{\pi}_k + \bar B^0 A^0 + \bar A^0 B^0 \right]^2,
\end{gather}
where $A_{\dot\alpha} A_{\dot\alpha} = 2 \tilde{a}^+ \tilde{a}^- +
\tilde{a}^0 \tilde{a}^0 + 2 \bar A^0 A^0$,  $B_{\dot\alpha} B_{\dot\alpha} = 2 \tilde{\pi}^+ \tilde{\pi}^- + \tilde{\pi}^0 \tilde{\pi}^0 + 2 \bar B^0 B^0 $ .

The complete set of the lightest spin-0 H-hadrons in the model
includes pNG states (pseudoscalar H-pions $\tilde\pi_k$ and scalar
complex H-diquarks/H-baryons $B^0$), their opposite-parity chiral partners
$\tilde a_k$ and $A^0$, and singlet H-mesons $\tilde\sigma$ and
$\tilde\eta$. These H-hadrons are listed in Table
\ref{tab:H-hadrons} along with their quantum numbers and associated
H-quark currents. Note that the total Lagrangian of the model given
by \eqref{eq:lp}, \eqref{eq:DQ}, \eqref{eq:LM}, and
\eqref{eq:U} is invariant under a global transformation
\begin{gather}
Q' = e^{\frac{i}2\xi} Q,
\qquad (A^0)'=e^{i\xi} A^0,
\qquad (B^0)'=e^{i\xi} B^0
\end{gather}
or equivalently the Lagrangian given by \eqref{eq:ctq}, \eqref{eq:LM}, and \eqref{eq:U} in terms of the quartet field $P_L$ and the
antisymmetric field $M$ is invariant under a transformation
\begin{gather} \label{eq:UHB2}
P_L' = e^{\frac{i}2\xi \Sigma_4} P_L,
\qquad M'=e^{\frac{i}2\xi \Sigma_4} M e^{\frac{i}2\xi \Sigma_4^T},
\qquad \Sigma_4 = \frac1{2\sqrt2} \begin{pmatrix} 1 & 0 \\ 0 & -1 \end{pmatrix},
\end{gather}
where $\Sigma_4$ is a generator of $Sp(4) \subset SU(4)$. The EW symmetry, which is spanned by the generators $\Sigma_k$, $k=1,\,2,\,3$
defined by \eqref{eq:Sigma123}, does not break the symmetry \eqref{eq:UHB2}, since the generator $\Sigma_4$ commutes with $\Sigma_k$.
This additional global $U(1)_{HB}$ symmetry \eqref{eq:UHB2} allows us to introduce a conserved H-baryon number, which makes the lightest H-diquark stable.
We remind that the model contains the elementary Higgs field which is not a pNG state. There is, however, a scenario with a composite
Higgs boson having also a new strongly coupled sector with the  symmetry breaking pattern $SU(4)\to Sp(4)$ \cite{Bizot:2016}.

\begin{table}
\caption{\label{tab:H-hadrons}Quantum numbers of the lightest (pseudo)scalar H-hadrons and the corresponding H-quark currents in $SU(2)_{HC}$ model.
$\tilde G$ denotes hyper-$G$-parity of a state (see Section \ref{sec:physlagr}). $\tilde B$ is the H-baryon number. $Q_\text{em}$ is the electric charge.
$T$ is the weak isospin. Hyperbaryons do not carry intrinsic $C$- and $HG$-parities, since the charge conjugation reverses the sign of the H-baryon number.}
\begin{tabular}{ccccccccc}
\hline
state & $$ & H-quark current & $$ & $T^{\tilde G}(J^{PC})$ & $$ & $\tilde B$ & $$ & $Q_\text{em}$ \\
\hline
$\tilde\sigma$ & $$ & $\bar Q Q$ & $$ & $0^+(0^{++})$ & $$ & 0 & $$ & 0 \\
$\tilde\eta$ & $$ & $i \bar Q \gamma_5 Q$ & $$ & $0^+(0^{-+})$ & $$ & 0 & $$ & 0 \\
$\tilde a_k$ & $$ & $\bar Q \tau_k Q$ & $$ & $1^-(0^{++})$ & $$ & 0 & $$ & $\pm 1$, 0 \\
$\tilde\pi_k$ & $$ & $i \bar Q \gamma_5 \tau_k Q$ & $$ & $1^-(0^{-+})$ & $$ & 0 & $$ & $\pm 1$, 0 \\
$A^0$ & $$ & $\bar Q_{a\underline a}{}^{C} \epsilon_{ab} \epsilon_{\underline a \underline b} Q_{b\underline b}$ & $$ & $0^{\hphantom{+}}(0^{-\hphantom{+}})$ & $$ & 1 & $$ & 0 \\
$B^0$ & $$ & $i \bar Q_{a\underline a}{}^{C} \epsilon_{ab} \epsilon_{\underline a \underline b} \gamma_5 Q_{b\underline b}$ & $$ & $0^{\hphantom{+}}(0^{+\hphantom{+}})$ & $$ & 1 & $$ & 0 \\
\hline
\end{tabular}
\end{table}

\section{\label{sec:physlagr}Physical Lagrangian of the minimal model}

Now, we represent the part of physical Lagrangian which is relevant
for further analysis of the most interesting case with zero
hypercharge (stable H-pion scenario). The H-quark interactions
with the EW bosons are vectorlike, and the corresponding Lagrangian
follows from (\ref{eq:DQ}):
\begin{align}\label{3.1}
L(Q,G)={}&\frac{1}{\sqrt{2}}g_W\bar{U}\gamma^{\mu}D
W^+_{\mu}+\frac{1}{\sqrt{2}}g_W\bar{D}\gamma^{\mu}U W^-_{\mu}\notag\\
&+\frac{1}{2}g_W(\bar{U}\gamma^{\mu}U-\bar{D}\gamma^{\mu}D)(c_W
Z_{\mu}+s_W A_{\mu}).
\end{align}
Here $c_W$ and $s_W$ denote cosine and sine of the Weinberg angle.
Interactions of (pseudo)scalars with photons and intermediate bosons
are described by the following Lagrangians:
\begin{gather}\label{eq:SG1}
    {L}(\tilde\sigma/H,G)
    = \frac18 \left[ 2 g_W^2 W_\mu^+ W^\mu_- + ( g_B^2 + g_W^2 ) Z_\mu Z^\mu \right] (\cos\theta_s H - \sin\theta_s \sigma)^2,
\end{gather}
\begin{gather}\label{eq:SG2}
 L(\tilde\pi/\tilde a, G) = \left[  i g_W W_+^\mu \left( \tilde\pi^0  \tilde\pi^-_{,\mu}- \tilde\pi^-
\tilde\pi^0_{,\mu} \right) + \text{h.c.} \right] + i g_W ( c_W
Z^\mu - s_W A^\mu ) (\tilde\pi^- \tilde\pi^+_{,\mu}-\tilde\pi^+
\tilde\pi^-_{,\mu})
\notag\\
 + g_W^2 \tilde \pi^+ \tilde\pi^- ( c_W Z^\mu - s_W A^\mu )^2
\notag\\
 - g_W^2 \tilde\pi^0 ( c_W Z^\mu - s_W A^\mu ) \left( \tilde\pi^+ W^-_\mu + \tilde\pi^- W^+_\mu \right)
\notag\\
 -\frac12 g_W^2 \left( \tilde\pi_+^2 W^-_\mu W_-^\mu  + \tilde\pi_-^2 W^+_\mu W_+^\mu \right) + g_W^2
 \left( \tilde\pi_0^2+\tilde\pi^- \tilde\pi^+ \right) W^+_\mu W_-^\mu +(\tilde\pi \to \tilde a)  .
\end{gather}
In the above Lagrangian $L(\tilde\pi/\tilde a, G)$ the last term means that the interactions of the triplet
scalar H-mesons $\tilde a$ have the same couplings and vertices as the interactions of the H-pions.

The scalar and pseudoscalar fields $\tilde\sigma$, $\tilde\pi$, $H$ interaction with the H-quarks is
described by the Lagrangian which directly follows from
\eqref{eq:lp}:
\begin{align}\label{3.3}
L(Q,\tilde\sigma,H)={}&-\kappa (c_{\theta} \tilde{\sigma}+s_{\theta}
H)(\bar{U}U+\bar{D}D)+i\sqrt{2}\kappa \tilde{\pi}^+\bar{U}\gamma_5
D\notag\\&+i\sqrt{2}\kappa \tilde{\pi}^-\bar{D}\gamma_5 U+i\kappa
\tilde{\pi}^0(\bar{U}\gamma_5 U-\bar{D}\gamma_5 D),
\end{align}
where $c_\theta= \cos\theta_s$ and $s_\theta= \sin\theta_s$.
There is a specific symmetry of the minimal hypercolor model leading
to some phenomenological consequences. At the fundamental level, the Lagrangian of the current H-quarks \eqref{2.10} is invariant under modified charge
conjugation of the H-quark fields (hyper-$G$-parity, $HG$-parity) which
is defined as follows:
\begin{equation}\label{3.4}
    (Q_{a\underline a})^{HG}= \epsilon_{ab}\epsilon_{\underline a \underline b}
    Q^{C}_{b\underline b},
\end{equation}
where $C$ is the charge conjugation, $a,b$ are isotopic indices, and
$\underline a$, $\underline b$ are hypercolor indices (it is the same notation
as in the Section \ref{sec:2}). To prove the statement, we use
(\ref{2.6}) and the properties of bilinear forms with respect to
the ordinary charge conjugation
\begin{align}\label{3.6}
    \bar{Q}^C_{a\underline a}Q^C_{b\underline b}=&\bar{Q}_{b\underline b}Q_{a\underline a},\,\,\,\bar{Q}^C_{a\underline a}\gamma_5Q^C_{b\underline b}=
    \bar{Q}_{b\underline b}\gamma_5Q_{a\underline a},\notag\\
    &\bar{Q}^C_{a\underline a}\gamma_{\mu}Q^C_{b\underline b}=-\bar{Q}_{b\underline b}\gamma_{\mu}Q_{a\underline a}.
\end{align}
By straightforward calculations one can check that the Lagrangian
\eqref{2.10} is invariant under the transformation \eqref{3.4}, since
the H-gluon $T_{\mu}$ and the SM fields are not transformed.
To analyze transformation properties of the
$\tilde{\pi}\bar{Q}Q$ effective vertex in more detail, we use
(\ref{3.4})  and (\ref{3.6}) and have
\begin{align}\label{3.7}
    (\bar{Q}_{a\underline a}\gamma_5 \tau^k_{ab}\tilde{\pi}_k
    Q_{b\underline a})^{HG}&=\bar{Q}^{C}_{a\underline a}\epsilon_{ab}\epsilon_{\underline a\underline b}\gamma_5
    \tau^k_{bc}\tilde{\pi}^{HG}_k\epsilon_{cd}\epsilon_{\underline b\underline c}Q^C_{d\underline c}\notag\\
    &=-\bar{Q}^C_{a\underline a}\gamma_5\tau^{*k}_{ad}\tilde{\pi}^{HG}_k Q^C_{d\underline a} =-\bar{Q}_{a\underline a}\gamma_5 \tau^k_{ab}\tilde{\pi}_k^{HG}
    Q_{b\underline a}.
\end{align}

So, the invariance condition results in the transformation
$\tilde{\pi}^{HG}_k=-\tilde{\pi}_k$, that is, $\tilde{\pi}$ is odd,
while the SM fields are even under modified charge conjugation
\eqref{3.4}. This is a special case of the treatment of general
vectorlike HC models in Refs.
\cite{2010PhRvD..82k1701B,Antipin:2015xia}. It is observed in
\cite{2010PhRvD..82k1701B} that $HG$-parity is a good quantum number
of the theory and all SM particles are $HG$-even. Thus, $HG$-odd
$\tilde{\pi}$ has not decay modes with only SM particles in the
final states. In the model under consideration decay
channels of type $\tilde{\pi}^{\pm}\to \tilde{\pi}^0 X^{\pm}$ are
allowed due to $HG$-parity conservation.

It is important, all restrictions on the oblique corrections are
fulfilled in this variant of hypercolor. If the hypercharge is zero and $h$--$\tilde{\sigma}$ mixing is absent, then $T$-parameter is
equal to zero. If, however, we consider a HC scenario with a
non-zero hypercharge and mixing, a constraint for the $T$ parameter
value emerges (see \cite{TC12ou,DMou1}). Then the
$h$--$\tilde{\sigma}$ mixing angle should be sufficiently small to
avoid problems with the PT parameters and the measured properties of
the SM Higgs boson.

\section{Low-energy signature of the model}

In this section, we consider briefly main phenomenological
consequences of the minimal model for the case of zero hypercharge.
In spite of a simple structure and minimal particle content, the
model can manifest a rich phenomenology and interesting signature in
collider physics. Here we consider processes with the H-sigma
$(\tilde{\sigma})$ and H-pions $(\tilde{\pi})$.
It is supposed that
these states are the lowest ones in the model (see, however, the
results of lattice calculations in \cite{Arthur:2016ozw}).
Indeed, the claim that pNG states are the lightest in the mass spectrum is based on the hypothesis of a hierarchy of H-physics scales.
In other words, we suppose that other (not pNG) possible H-hadrons including vector H-mesons are heavier than the pNG bosons. Namely,
the explicit $SU(4)$ symmetry breaking  is considered  as a small perturbation in comparison with the dynamical symmetry breaking
in analogy with the orthodox QCD, where the scale of chiral symmetry breaking is much larger than the light quark masses.
 From our previous analysis of the parameter space, it follows that the masses
of H-mesons are of the order of $10^2\text{--}10^3\, \mbox{GeV}$.
Thus, the low-energy pNG states of the minimal model can be accessed
at the LHC and future linear collider.

Channels of H-pion production and decay are described by the model Lagrangian (see the previous Section).
At the LHC these pNG states most effectively occur in two ways: in vector boson fusion (VBF) reaction
$pp\to V^* V^{'*}\to \tilde{\pi} \tilde{\pi}'$, where
$V = W,Z,\gamma$, or in the $s$-channel of $q {\bar q}'$- or $q \bar q$-fusion---Drell--Yan type (DYT) process, $pp\to V^{'*}\to
\tilde{\pi}\tilde{\pi}$. Corresponding Feynman diagrams can be found in
\cite{TC12ou}. There is also an analog of usual associated production where H-pion pair is produced together with vector boson, $pp\to V^{'*}\to
V \tilde{\pi}\tilde{\pi}$. Its cross section, is
somewhat suppressed compared with the DYT reaction by extra factor $g_W^2$. The channel, however, has a specific set of final states (see below).

As to VBF and DYT mechanisms, their contributions to the
cross section of H-pion pair production strongly depend on the
invariant H-pion mass, kinematic cuts for final states, quark pdf's,
combinatorial factors and $q \to Vq'$ splitting functions at high
energies. Of course, NLO and NNLO corrections for these channels
should be different and can be important---as is the case for Higgs
production at the LHC \cite{DenDitAP,CicDenDit,FGT,MMZ}. A detailed analysis of LO cross sections and NLO corrections is
beyond the scope of the paper.

It seems that the VBF production of H-pions is suppressed, in
particular, by an additional $g_W^4$, and Drell--Yan type process dominates
(see Ref.~\cite{Sundrum}). The situation is, however, more
complicated due to the above mentioned factors, and in the TeV region
$VV'$-fusion cross section is very close to DYT or even larger
(see, for example, Ref.~\cite{MMZ}). Moreover, due to suitable $p_T$
cuts it is possible that, as it happens for the high mass ($\sim
\,\mbox{TeV}$) scalar boson production \cite{CicDenDit}, $s$-channel
$qq'$-fusion cross section should be comparatively small. Of
course, it is not the same process, nevertheless, enhancing factors
for the VBF are analogous---a lot of integrated partons with low $x$
and $p_T$ when vector boson splits off. Namely, due to integration
with quark splitting functions in the region of low partonic $p_T$,
VBF cross section can be increased by $\log^2(M/M_W)$, $M$ is an invariant mass of H-pion pair. Note also that large
resonance $s$-channel contribution into VBF production with
intermediate $\tilde{\sigma}$ is possible if $m_{\tilde{\sigma}}$ is close to
$2m_{\tilde{\pi}}$. This point should be considered separately.

VBF cross section of H-pion pair
production, as function of $qq'$ center-of-mass energy and H-pion
mass, was calculated in our paper \cite{TC12ou} and $\sigma_\text{VBF} (pp \to qq' \tilde{\pi}\tilde{\pi}) \approx (0.01\text{--}0.02)\,\mbox {pb}$ when $E_{cm} \sim 1\, \mbox{TeV}$
and H-pion mass is $200\text{--}300 \, \mbox{GeV}$.
We also estimate the DYT cross section in this region as approximately $0.03\text{--}0.05 \,\mbox {pb}$. Both of these cross sections decrease of about one order of magnitude
with the mass of $\tilde \pi$ increasing up to $\sim 500\text{--}700 \, \mbox {GeV}$.

Almost the same situation is observed for the hierarchy of Higgs production mechanisms \cite{Handbook}---associated Higgs production dominates
at $\sqrt{s} = 2\, \mbox{TeV}$,---but at higher energies, $\sqrt{s} \geq 4\, \mbox{TeV}$, the situation is reversed and VBF cross section exceeds associated production by almost a half.
In other words, behavior of these cross sections at high energies should be studied more carefully and it will be done in the next paper.

Estimated cross sections of H-pion production are small, so, to detect a signal, large statistics
and the background suppression are necessary. From this point of view, VBF reactions are more
perspective due to two hard tagging quark jets. Adding some reasonable cuts, for the rapidity to highlight
the central region of the reaction, $ |\eta| \leq 2.5$, and for final leptons,
$p_T \geq 100 \, \mbox{GeV}$, it is possible to separate leptons from $\tilde \pi^{\pm}$ decay.
These decays are also marked by large missed $p_T$ due to heavy stable neutral H-pions and neutrino.

 H-pion production in the process of annihilation $e^+e^-\to \gamma^*
,Z^*\to 2\tilde{\pi}; 4\tilde{\pi}$ is also possible through the
reactions of the type $Z^*\to\tilde{\pi}^+\tilde{\pi}^-$ and
$W^{\pm*}\to \tilde{\pi}^{\pm}\tilde{\pi}^0$. These processes have
transparent signature and can be studied at future linear colliders.
Note, some interesting features should be observed: production of
H-pions in associated process, $e^+e^- \to Z^* \to Z \tilde
\pi^{\pm} \tilde \pi^{\mp}$, $e^+e^- \to Z^* \to W \tilde \pi^{\pm}
\tilde \pi^0$, or via $VV'$-fusion. At the ILC Higgs boson production
cross sections demonstrate evolution with energy \cite{BorKa} which
is analogous to predicted for the LHC. In the H-pion production we
expect the same behavior of cross sections.

To analyze a final signature in the reactions above, note that due to $HG$-parity conservation (see the previous Section)
H-pions have no tree-level decay modes having in the final states
the SM particles only. The lowest order amplitudes which govern
decays of the type $\tilde{\pi}\to V_1V_2,\,V_1V_2V_3$, where
$V_a=\gamma,Z,W$, are described by triangle and box diagrams with
H-quarks loops. It can be easily checked that interference
contributions for the transition $\tilde{\pi}\to V_1V_2$ with $U$
and $D$ hyperquark loops cancel out each other. Since $M_U=M_D$,
this compensation is obvious due to the opposite charges,
$q_D=-q_U$, when $Y_1=0$. Analysis of the decay $\tilde{\pi}\to
V_1V_2V_3$ reveals compensation of box contributions. More exactly,
the diagrams with loop momenta circulating in opposite directions
cancel out each other. It is easy to prove that the compensation
results from generalized Furry's theorem \cite{1951PThPh...6..614N}.
To this end, we should use the following properties of the Dirac and
Pauli matrices:
\begin{align}\label{4.1}
   &\Tr\{\gamma_{\mu_1}\gamma_{\mu_2}....\gamma_{\mu_n}\}=\Tr\{\gamma_{\mu_n}....\gamma_{\mu_2}\gamma_{\mu_1}\},\notag\\
   &\Tr\{\tau_{a_1}\tau_{a_2}....\tau_{a_n}\}=(-1)^n
   \Tr\{\tau_{a_n}....\tau_{a_2}\tau_{a_1}\}.
\end{align}

The cancellation of amplitudes in the case of an even number of
final bosons is inherently isotopic---it results from the zero
H-quark hypercharge. If the number of final bosons is odd, such
cancellation follows from the charge parity conservation along with
the vectorlike structure of the H-quark EW interaction. As a
result, the H-pion fields are stable in the framework of the
vectorlike hypercolor model with zero hypercharge and degenerate
masses in the H-quark doublet $Q$ and triplet $\tilde{\pi}$. In the
previous section, it was demonstrated that this stability follows from
the presence of the discrete symmetry in the model.

Note, the H-quark masses remain degenerate, $M_U = M_D$, at the one
loop level. It can be easily checked that the self-energy
contributions into the mass renormalization are defined by
electroweak and H-pion loops. These terms are exactly the same for
the $U$ and $D$ quarks. However, this effect does not take place in
the case of the H-pion masses. The mass-splitting value of the
H-pion can be calculated by summing over self-energy diagrams.
Detailed analysis of the relevant amplitudes reveals that only EW
diagrams contribute into the mass-splitting $\Delta
m_{\tilde{\pi}}=m_{\tilde{\pi}^{\pm}}-m_{\tilde{\pi}^0}$, all strong
(H-quark) loops are canceled out. As a result we get
\begin{align}\label{4.2}
\Delta m_{\tilde{\pi}} = {}&\frac{G_F M_W^4}{2\sqrt{2} \pi^2 m_{\tilde{\pi}}} \Biggl[ \ln \frac{M_Z^2}{M_W^2} -\beta_Z^2 \ln \mu_Z+ \beta_W^2 \ln \mu_W \notag\\
&-\frac{4 \beta_Z^3}{\sqrt{\mu_Z}}\left ( \arctan \frac{2-\mu_Z}{2\sqrt{\mu_z}\beta_Z} +\arctan \frac{\sqrt{\mu_Z}}{2\beta_Z} \right)\notag\\
&+\frac{4 \beta_W^3}{\sqrt{\mu_W}}\left( \arctan
\frac{2-\mu_W}{2\sqrt{\mu_W}\beta_W} +\arctan
\frac{\sqrt{\mu_W}}{2\beta_W} \right) \Biggr],
\end{align}
where $\mu_V=M_V^2/m_{\tilde{\pi}}^2,\quad
\beta_V=\sqrt{1-\mu_V/4}$, and $G_F$ is Fermi's constant.
 For the H-pion masses in the interval $200\text{--}800 \, \mbox{GeV}$
from (\ref{4.2}) it follows that $\Delta m_{\tilde{\pi}} \approx
0.170\text{--}0.162 \, \mbox{GeV}$. Non-zero mass-splitting in the
H-pion triplet violates isotopic invariance. However, $HG$-parity
remains a conserved quantum number since it is induced by a discrete
symmetry rather then a continuous transformation in the space of
H-pion states. Thus, an account of higher order corrections does not
lead to destabilization of the neutral H-pion.

So, the analysis performed leads to the conclusion that the model
involves stable weakly interacting neutral H-pion. Then, production
of the H-pions at colliders manifests itself with some unique
signature of the final state---charge leptons and large missing energy.
The stable H-pion $\tilde{\pi}^0$ can be also considered as a
component of Dark Matter.

For the width of the charged H-pion decay in the strong channel we
get:
\begin{equation}\label{4.3}
\Gamma(\tilde{\pi}^{\pm}\to\tilde{\pi}^0
\pi^{\pm})=\frac{G^2_F}{\pi}f^2_{\pi}|U_{ud}|^2
m_{\tilde{\pi}}^{\pm} (\Delta
m_{\tilde{\pi}})^2\bar\lambda(m^2_{\pi^\pm},m_{\tilde{\pi}^0}
^2;m_{\tilde{\pi}^{\pm}}^2).
\end{equation}
Here $f_{\pi} =132 \,\mbox{MeV}$, $\pi^{\pm}$ is a standard pion and
\begin{equation}\label{4.4}
\bar\lambda(a,b;c)=\left[1-2\frac{a+b}{c}+\frac{(a-b)^2}{c^2}\right]^{1/2}.
\end{equation}

The H-pion decay width in the lepton channel is
\begin{equation}\label{4.5}
\Gamma(\tilde{\pi}^{\pm}\to\tilde{\pi}^0 l^{\pm}\nu_l)=\frac{G^2_F
    m_{\tilde{\pi}^{\pm}}^3}{24\pi^3}\int_{q^2_1}^{q^2_2}\bar\lambda(q^2,m_{\tilde{\pi}^0}^2;m_{\tilde{\pi}^{\pm}}^2)^{3/2}
\left(1-\frac{3m^2_l}{2q^2}+\frac{m^6_l}{2q^6}\right)\,dq^2,
\end{equation}
where $q^2_1=m^2_l$, $q^2_2=(\Delta m_{\tilde{\pi}})^2$, and $m_l$
is a lepton mass.

Now, using \eqref{4.3}, \eqref{4.5}, and the value $\Delta m_{\tilde{\pi}}$ from \eqref{4.2}, we estimate decay widths, lifetimes, and proper decay lengths in these channels as
follows $\footnote{In \cite{Beylin2017}, the designation of the decay channels is erroneously rearranged. Here we give the correct results.}$:
\begin{align}\label{4.6}
\Gamma(\tilde{\pi}^{\pm}\to\tilde{\pi}^0 l^{\pm}\nu_l)&=6\cdot
10^{-17}\,\mbox{GeV},\,\,\,\tau_{l}=1.1\cdot10^{-8}\,\mbox{sec},\,\,\,c\tau_{l}\approx 330\,\mbox{cm};\notag\\
\Gamma(\tilde{\pi}^{\pm}\to\tilde{\pi}^0 \pi^{\pm} )&=3\cdot10^{-15}\,\mbox{GeV},\,\,\,\tau_{\pi}=2.2\cdot10^{-10}\,\mbox{sec},\,\,\,c\tau_{\pi}\approx 6.6\,\mbox{cm}.
\end{align}

From these analysis it follows that main characteristic fingerprints of
H-pions at TeV scale in the VBF, DYT and associated production are:

\begin{enumerate}

\item $V^*V^* \to \tilde \pi \tilde \pi + jj$ --- two hard tagging jets, high $p_{T,mis}$ from two $\tilde \pi^0$, neutrino and a lepton (or two charged leptons) from $\tilde \pi^{\pm}$;

\item $V^* \to \tilde \pi \tilde \pi$ ---high $p_{T,mis}$ from two $\tilde \pi^0$ and $\nu_l$, and final one lepton, $l\bar l$ or $\pi^{\pm} \pi^{\pm}$ from pair of $\tilde \pi^{\pm}$;

\item $V^* \to V \tilde \pi \tilde \pi$ --- hadron jets (or $l \bar l$ or $l\nu_l$) from $W$ or $Z$ decays, high $p_{T,mis}$ from two $\tilde \pi^0$ and neutrino
(from $\tilde \pi^{\pm}$ and/or $W^{\pm}$); $l^+l^-$ - if there are two final charged H-pions or one charged H-pion and $W$,
tri-lepton signal from $W \tilde \pi^{\pm} \tilde \pi^{\pm}$ final state.
\end{enumerate}

As to the production of a single scalar H-sigma $\tilde{\sigma}$ at
the LHC and ILC, it is strongly suppressed reaction at the tree
level due to the small $\tilde{\sigma}$--$h$ mixing. More exactly,
the tree-level $\tilde{\sigma}$ production is suppressed with
respect to the Higgs production by $\sin^2 \theta_s$, where $\theta_s$
is a mixing angle.

At the one-loop level both single and double H-sigma production
occur in the processes of type $V^{*}V'^{*}\to \tilde{\sigma},
\,2\tilde{\sigma}$ and/or $V^{*}\to \Delta \to
V^{'}\tilde{\sigma},\,2\tilde{\sigma}$, where $V^{*}$ and $V^{'}$
are vector bosons in the intermediate and final states, $\Delta$
denotes a H-quark triangle loop.

Decays of the type $\tilde{\sigma}\to V_1V_2$, where
$V_{1,2}=\gamma,Z,W$, proceed through H-quark and H-pion loops.
Dominant decay channels of H-sigma are  $\tilde{\sigma}\to
\tilde{\pi}^0 \tilde{\pi}^0,\,\tilde{\pi}^+\tilde{\pi}^-$, which
take place at the tree level and provide large decay width for
$m_{\tilde{\sigma}}\geqslant 2m_{\tilde{\pi}}$. The width is mostly
defined by the coupling $\lambda_{11}$ in the limit of small mixing:
\begin{equation}\label{4.7}
\Gamma(\tilde{\sigma}\to\tilde{\pi}\tilde{\pi})=\frac{3u^2\lambda^2_{11}}{8\pi
    m_{\tilde{\sigma}}}\left(1-\frac{4m^2_{\tilde{\pi}}}{m^2_{\tilde{\sigma}}}\right).
\end{equation}
Using the previous parametric analysis in \cite{TC12ou} concerning
the value $\lambda_{11}$ ($\lambda_{HC}$ in \cite{TC12ou}) and $u$,
from (\ref{4.7}) one can get
$\Gamma(\tilde{\sigma}\to\tilde{\pi}\tilde{\pi})\gtrsim
10\,\mbox{GeV}$ when $m_{\tilde{\sigma}}\gtrsim 2m_{\tilde{\pi}}$.

As it was noted above, the small mixing $h$--$\tilde{\sigma}$ in
conformal approximation leads to the relation
$m_{\tilde{\sigma}}\approx\sqrt{3}m_{\tilde{\pi}}$ and all
tree-level decay widths are proportional to the square value of the
$\tilde{\sigma}$--$h$ mixing angle $\theta_s$. Corresponding decay
widths are as follows:
\begin{align}\label{4.7a}
\Gamma(\tilde{\sigma}\to f\bar{f})=&\frac{g_W^2 \sin^2 \theta_s}{32\pi}m_{\tilde{\sigma}}\frac{m^2_f}{M^2_W}(1-4\frac{m^2_f}{m^2_{\tilde{\sigma}}})^{3/2},\notag\\
\Gamma(\tilde{\sigma}\to ZZ)=&\frac{g_W^2 \sin^2 \theta_s}{16\pi
c^2_W}\frac{M^2_Z}{m_{\tilde{\sigma}}}(1-4\frac{m^2_Z}{m^2_{\tilde{\sigma}}})^{1/2}[1+\frac{(m^2_{\tilde{\sigma}}-2M^2_Z)^2}{8M^4_Z}],\notag\\
\Gamma(\tilde{\sigma}\to W^+W^-)=&\frac{g_W^2 \sin^2
\theta_s}{8\pi}\frac{M^2_W}{m_{\tilde{\sigma}}}(1-4\frac{m^2_W}{m^2_{\tilde{\sigma}}})^{1/2}
[1+\frac{(m^2_{\tilde{\sigma}}-2M^2_W)^2}{8M^4_W}].
\end{align}
In (\ref{4.7a}) $m_f$ is a mass of standard fermion $f$ and
$c_W=\cos \theta_W$. In the limit of zero mixing we should consider
the loop-level decay channels. Here, we consider the decay channel
$\tilde{\sigma}\to \gamma\gamma$ which proceeds mainly through
H-quark and H-pion loops. The width can be written in the form
\begin{equation}\label{4.8}
\Gamma(\tilde{\sigma}\to
\gamma\gamma)=\frac{\alpha^2m_{\tilde{\sigma}}}{16\pi^3
    } |F_Q+F_{\tilde{\pi}}+F_{\tilde{a}}+F_W+F_\text{top}|^2,
\end{equation}
where the contributions of H-quarks, $F_Q$, H-pions,
$F_{\tilde{\pi}}$, $W$-bosons, $F_W$, and top-quarks, $F_\text{top}$, are
defined by the following expressions:
\begin{align}\label{4.9}
F_Q&=-2\kappa \frac{M_Q}{m_{\tilde{\sigma}}}[1+(1-\tau^{-1}_Q)f(\tau_Q)], \notag \\
F_{\tilde{\pi}}&=\frac{g_{\tilde{\pi}\tilde{\sigma}} }{m_{\tilde{\sigma}}}[1-\tau^{-1}_{\tilde{\pi}}f(\tau_{\tilde{\pi}})],\,\,\,g_{\tilde{\pi}\tilde{\sigma}}\approx u\lambda_{11}, \notag \\
F_{\tilde{a}}&=\frac{g_{\tilde{a}\tilde{\sigma}}}{m_{\tilde{\sigma}}}[1-\tau^{-1}_{\tilde{a}}f(\tau_{\tilde{a}})],\,\,\,g_{\tilde{a}\tilde{\sigma}}\approx u\lambda_{12}, \notag \\
F_W&=-\frac{g_W\sin\theta_s m_{\tilde{\sigma}}}{8M_W}
[2+3 \tau_W^{-1} + 3 \tau_W^{-1}(2-\tau_W^{-1}) f(\tau_W)], \notag \\
F_\text{top}&=\frac{4}{3}\frac{g_W\sin \theta_s M^2_t}{m_{\tilde{\sigma}}M_W}
[1+(1-\tau^{-1}_t)f(\tau_t)],
\end{align}
and
\begin{align}\label{4.10}
f(\tau)=&\arcsin^2\sqrt{\tau},\,\,\,\tau<1,\notag\\
f(\tau)=&-\frac{1}{4}\Biggl[\ln\frac{1+\sqrt{1-\tau^{-1}}}{1-\sqrt{1-\tau^{-1}}}-i\pi\Biggr]^2,\,\,\,\tau
>1.
\end{align}
Non-zero $\tilde{\sigma}\text{--}h$ mixing influences the width via
$W$- and $t$-quark loops, their amplitudes being proportional to
$\sin \theta_s$. Using the analysis of the model parameter space in
\cite{TC12ou}, we get an estimation $\Gamma(\tilde{\sigma}\to
\gamma\gamma) \approx 5\text{--}10 \,\, \mbox{MeV}$. To calculate
the $\tilde{\sigma}$ production in full processes $p\bar{p}\to
\tilde\sigma \to all$, a corresponding program which integrates
partonic cross sections with the quark distribution functions should
be used. Instead, we give an approximate evaluation of this cross
section for the sub-process of vector boson fusion
$VV\to\tilde\sigma(s)\to all$, where $V=\gamma, Z, W$. Namely, the
cross section can be calculated with sufficient accuracy using a
simple formula in the framework of factorization method
\cite{2013CEJPh..11..182K1}:
\begin{equation}\label{4.11}
\sigma(VV\to\tilde{\sigma}(s))=\frac{16\pi^2\Gamma(\tilde{\sigma}(s)\to
VV)}{9\sqrt{s}\,\bar{\lambda}^2(M_V^2,M_V^2;s)}\,\rho_{\tilde{\sigma}}(s),
\end{equation}
where $\tilde{\sigma}(s)$ is $\tilde{\sigma}$ in the intermediate
state with energy $\sqrt{s}$ and $\Gamma(\tilde{\sigma}(s)\to VV)$
is a partial width. The probability density $\rho_{\tilde{\sigma}}(s)$
is defined by the following expression:
\begin{equation}\label{4.12}
\rho_{\tilde{\sigma}}(s)=\frac{1}{\pi}\frac{\sqrt{s}\,\Gamma_{\tilde{\sigma}}(s)}{(s-M^2_{\tilde{\sigma}})^2+s\,\Gamma^2_{\tilde{\sigma}}(s)},
\end{equation}
where $\Gamma_{\tilde{\sigma}}(s)$ is the total width of the
$\tilde{\sigma}$ with a mass squared equal to $s$. Exclusive
cross section at peak energy region $\sqrt{s}=M_{\tilde{\sigma}}$
can be found by the change in the numerator of the expression
(\ref{4.12}) $\Gamma_{\tilde{\sigma}}\,\to\,\Gamma(\tilde{\sigma}\to
V^{'} V^{'})=\Gamma_{\tilde{\sigma}}\cdot Br(\tilde{\sigma}\to V^{'}
V^{'})$:
\begin{align}\label{4.13}
\sigma(VV\to\tilde{\sigma}\to V^{'}
V^{'})=&\frac{16\pi}{9}\frac{Br(\tilde{\sigma}\to
VV)Br(\tilde{\sigma}\to V^{'}
V^{'})}{m^2_{\tilde{\sigma}}(1-4M^2_V/m^2_{\tilde{\sigma}})}\notag\\&\approx\frac{16\pi}{9
m^2_{\tilde{\sigma}}}\cdot Br(\tilde{\sigma}\to
VV)Br(\tilde{\sigma}\to V^{'}V^{'}).
\end{align}
So, the cross section at $m^2_{\tilde{\sigma}}\gg M^2_V$ is fully
defined by branchings of sigma decay and $m_{\tilde{\sigma}}$. When
$2m_{\tilde{\pi}}>m_{\tilde{\sigma}}$ dominant decay channels are
$\tilde{\sigma}\to WW, ZZ$, which lead to a narrow peak
$(\Gamma\lesssim 10\text{--}100\, \mbox{MeV}$). However, here we have the
cross section of the sub-process and do not take into account the
distribution function. Moreover, we should also to average
cross section over energy resolution. Both these factors reduce
significantly the value of cross section. When
$2m_{\tilde{\pi}}<m_{\tilde{\sigma}}$ dominant decay channel is
$\tilde{\sigma}\to \tilde{\pi}\tilde{\pi}$ which leads to a wide
peak $(\Gamma\sim 10\, \mbox{GeV})$. In this case
$Br(\tilde{\sigma}\to VV)$ is small and we get very small
cross section. Thus, the main signature of the H-sigma production
and decay is a wide peak at $2m_{\tilde{\pi}}<m_{\tilde{\sigma}}$,
mostly caused by the strong possible decay $\tilde{\sigma}\to
2\tilde{\pi}$ along with weak signals caused by two-photon, lepton,
and quark-jet final states (from $WW$, $ZZ$, and standard
$\pi^{\pm}$ channels). There is also specific decay mode with two
stable neutral H-pions as products of $\tilde{\sigma}$ decay---this
manifests itself in a large missing energy together with charged
leptons in the final states. As it was shown in the end of Section
III, due to the global $U(1)_{HB}$ symmetry the lightest H-diquark is stable.
Then, from the physical Lagrangian it follows that the other
H-diquark can decay to the stable one and something else. So, there
is a possibility to construct the Dark Matter from two types of
particles: stable neutral H-pion and the lightest scalar (or
pseudoscalar) H-diquark with conserved H-baryon number. Detailed
consideration of the two-component scenario of the DM depends on the
variety of model parameters, mass splitting between the pNG states
and agreement with the data on the DM relic. We add that the
suggested DM model does not contain (stable) H-baryon carrying the
EW charge (see, for example, Ref.~\cite{Cline:2016nab}), so there
are no strong constraints for the DM relic in the case. The study is
in progress now and results will be presented in the next paper.

As to $A^0, \, B^0$ production at the colliders, these particles can be produced  only by intermediate pNG states,
$\tilde a_a, \, \tilde \eta$ and the Higgs boson,  $h$, or $ \tilde \sigma$. At the tree level these channels are suppressed by the mixing angle.
They also can originate from loops with the participation of pNG.

\section{Conclusion}
The analysis performed demonstrates some unique features of the
simplest minimal HC model with two generations of H-quarks and
$SU(2)_{HC}$ as the H-confinement group. This scenario makes it
possible to construct vectorlike interaction, starting from chiral
non-symmetric H-quark set of fields. In the simplest case of
two-flavor scenario the set of pNG bosons, (pseudo)scalar H-mesons
and H-baryons (H-diquarks), arises, which provides the rich
phenomenology. The neutral H-pion $\tilde{\pi}^0$ is stable when
$Y_1=0$ due to hyper-$G$-parity conservation, so specific decay channels
for $\tilde{\pi}^{\pm}$ and $\tilde{\sigma}$ with a large missing
energy open. Moreover, analysis of the production and decays of
(pseudo)scalar states, H-pion and H-sigma, allows to distinguish
between scenarios with zero and non-zero H-quarks hypercharge
\cite{TC12ou}. At the same time, the model predicts a strong signal
with large missing energy in the case
$2m_{\tilde{\pi}}<m_{\tilde{\sigma}}$ or weak signal with two-vector
final states in the opposite case. The presence of non-anomal global
symmetry $U(1)_{HB}$ in the model leads to the conservation of H-baryon
charge. This, in turn, manifests itself in the presence of the stable
H-baryon complex field $B^0$. Note, the H-baryon
state $A^0$ can be stable also when $M_{A^0}<M_{B^0}$. This possibility will be studied separately.

The minimal model under consideration has some phenomenological
features which can be verified both at collider experiments and by astrophysical
observations. An interesting consequence of the model structure is a
possible interpretation of the stable neutral H-pions and H-baryons
as particles of DM. So, the model with stable neutral fields gives
the possibility to construct two-component DM. In this work we concentrated mainly on the methodological aspects of the model.
To make complete phenomenological analysis, we
should consider astrophysical applications and take into account all
experimental restrictions on new physics.

\textbf{Conflict of Interests}\\The authors declare that there is no
conflict of interests regarding the publication of this paper.

\textbf{Acknowledgments.}\\The work of V.~B. and V.~K. was supported
by the grant \, 213.01-2014/013-BG provided by Southern Federal
University.


\end{document}